# Chapter 17

# Vulnerabilities of Smart Grid State Estimation against False Data Injection Attack

Adnan Anwar

School of Engineering & Information Technology
The University of New South Wales,
Canberra, ACT 2600, Australia.
Email: adnan.anwar@ adfa.edu.au

Abdun Naser Mahmood

School of Engineering & Information Technology
The University of New South Wales,
Canberra, ACT 2600, Australia.
Email: abdun.mahmood@adfa.edu.au

## Abstract

In recent years, Information Security has become a notable issue in the energy sector. After the invention of 'The Stuxnet worm' [1] in 2010, data integrity, privacy and confidentiality has received significant importance in the real-time operation of the control centres. New methods and frameworks are being developed to protect the National Critical Infrastructures like- energy sector. In the recent literatures, it has been shown that the key real-time operational tools (e.g., State Estimator) of any Energy Management System (EMS) are vulnerable to Cyber Attacks. In this chapter, one such cyber attack named 'False Data Injection Attack' is discussed. A literature review with a case study is considered to explain the characteristics and significance of such data integrity attacks.

**Keywords:** State Estimation, False Data Injection Attack, Smart Grid, Cyber Security, Data Integrity Attack.

## 17.1. Introduction

Power system State Estimation has been widely used at the utility control centres to know the system status during the power system operation. In order to ensure the stability and reliability of the power system, network operator monitors and controls the system states which are obtained from the state estimation processor. Generally, State Estimator provides an estimation of the data for all measured and unmeasured quantities. This advanced tool also filters out the measurement errors and noises and suppresses bad data. With the development of the power system research and engineering, the modern State Estimation programs have advanced capabilities which have enhanced the computational performances, as well as, accuracy. However, the challenges of accurate and efficient State Estimation programs have increased more because of the recent cyber attacks in the energy system infrastructure.

In a recent report of the 'Industrial Control Systems Cyber Emergency Response Team (ICS-CERT)', it has been mentioned that 198 cyber incidents happened in the financial year 2012 among which 41% happened within 'Energy sector'. In the first half of the financial year 2013, 200 incidents happened across all sectors of the critical infrastructure among which the highest attack (111 in incidents) happened in the energy sector (53%) as shown in Fig.17.1 [2].

From the last few years, distributed energy resources and storage devices are widely used which have changed the power flow patterns of the grid [3]. These renewable sources have intermittent nature and most of the time they are not dispatchable. Therefore, demand response has been a crucial issue in a smart grid environment. To face the challenges, Advanced Metering Infrastructures (AMI) which is equipped with smart meters, may play a significant role. It is obvious that the use of smart meters and advanced communication network has helped the utility operators to implement the SCADA controls more easily; however, the communication system of the cyber-physical smart grid has been more vulnerable in terms of cyber attack which may affect the communication network. These types of cyber related crimes may have devastating impact on the physical power grid including operational failures and loss of synchronization of different critical equipments of a power grid. Moreover, a large scale blackout may occur due to a cyber attack in a smart grid.

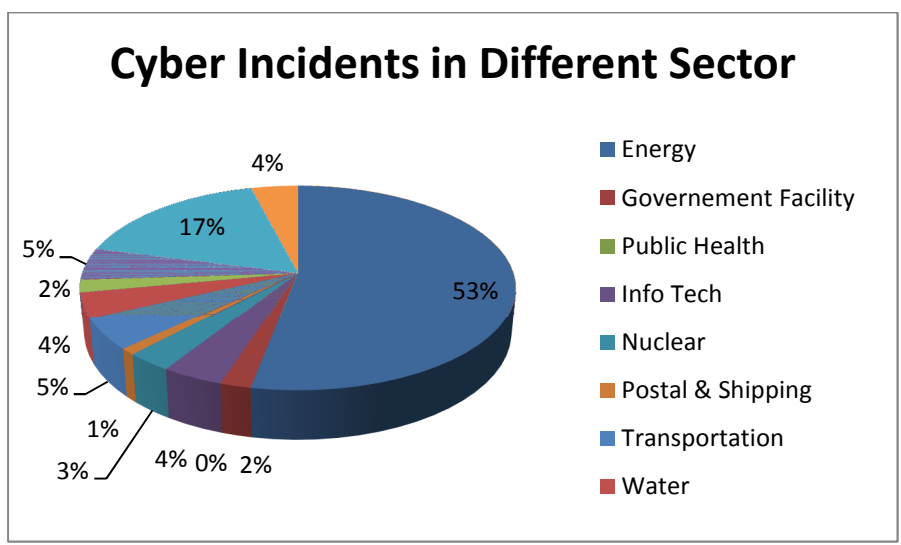

Fig.17.1: Cyber Incidents in Different Sector [2]

In a smart grid environment, energy system control sector needs advanced communication means among different parts of the network which increases the use of commercial off-the-shelf technologies. As a result, the cyber security issues arise. State Estimator, which is one key operational tool in the Energy Management System, is also very vulnerable to cyber attacks. Due to any pre-planned cyber attack in any State Estimation programs, bad data detectors may not be able to identify the possible threats which are attacked by any intruder. As a result, State Estimation programs will provide wrong information to the system operator. Based on the wrong estimation, operator may take misleading operational decision resulting a vital problem in the stable operation of power system. Therefore, advanced intrusion detection algorithms are desirable.

The organization of this chapter is as follows: A brief overview of traditional State Estimation is discussed in Section 17.2. The problem formulation with solution methodologies including Bad Data Detection techniques are also discussed in that section. The overview of Sate Estimation in Distribution System and Smart Grid is discussed in Section 17.3 and 17.4 respectively. A case study is illustrated to describe the vulnerabilities of the Smart Grid State Estimation in Section 17.5. A recent review on False Data Injection attack on State Estimator is discussed in Section 17.6. Finally, the chapter is concluded with brief remarks. This chapter intends to be a comprehensive reference in the field of cyber security of smart grid infrastructure.

## 17.2. Power System State Estimation

State Estimation is one of the most traditional power system analysis tools for reliable monitoring and control of Energy Management System. One early revolutionary work based on power system static state estimation was proposed in [4] and till then a significant number of research works have been conducted on this imperative issue. Although the traditional State Estimation has a long history in power transmission level, this powerful network analysis tool needs more attention in the low-voltage power distribution level [5], especially, when the grid adopts more communication infrastructures (i.e., Advanced Metering Infrastructure, Phasor Measurement Units, etc.) and Distributed Energy Resources (DERs). As mentioned earlier, the evolution of the Smart Grid State Estimation and the vulnerabilities of this estimation tool in terms of Data Integrity Attack will be discussed in the following section; this section will provide a brief overview of the significance of the traditional Static State Estimation and different methodologies and techniques involved with it. In this section, the importance of Bad Data identification and well-established procedures of detecting bad data are also discussed.

Basically, State Estimation is a procedure which is used to determine the most approximate solution of the system states by analyzing the measured sensor values and the equivalent calculated values. In power system theories and applications, the term 'state estimator' implies a computer program for calculating the system states based on the measured data at different nodes of the network and the laws of electric power networks which explains the behavior of the physical network model. The purpose of the state estimation is to estimate the unmeasured variables, improve overall efficiency and to detect the bad measurement.

Generally, the states in a power system are the complex voltage magnitude and the angles of each bus. If the state vector is **x**, then the state vector for an 'n' bus system will be:

$$\mathbf{x} = [\ \delta_2\ \delta_3\ \ldots\ldots \delta_n\ \ V_1\ V_2 V_3\ \ldots\ldots V_n]^T \qquad (17.1)$$

where, $\delta_i$ indicates the phase angles and $V_i$ the voltage magnitude at the *i*-th bus. It is interesting to note that, the dimension of the state vector is (2n-1)×1 as the phase angle at the reference bus is considered to be known which is generally assumed 0 rad. Although, the bus voltage magnitudes and angles are used in practice, current magnitudes-angles and power flows are also considered as state variables in some cases. At the first stage of the state estimation, measurement data are obtained from the Remote Terminal Units (RTUs) which are equipped with sensors. These measurements include voltage magnitudes, bus injections and both real and reactive power flows through different components of the network. However, the measurement data may be noisy and corrupted which increase the risk of direct use of these data. If the system states are known, then it is expected that from the laws of electric network (say, Kirchhoff's Current Law or Kirchhoff's Voltage Law) it is possible to calculate the power flow pattern of the network. However, it is not possible to directly measure the system states which motivate to develop and improve the methodologies related to State Estimations. As a result, at the second stage of state estimation, functions of state variables are used to calculate the expected values of the measurement data. Finally, any established method is employed to calculate the state variables from the measurement values and the calculated values. One such widely adopted method is Weighted Least Square (WLS) Method. After estimating the system states, the Bad Data Detection program is perform to identify the corrupted data. All of the steps of the state estimation are described briefly in following sub sections.

### 17.2.1. System Model of Measurement Data

Consider a measurement vector z for an n-bus system, where, $z \in R^{M \times 1}, M > (2n-1) \times 1$. Therefore, z should be:

$$\mathbf{z} = \begin{bmatrix} z_1 \\ z_2 \\ \vdots \\ z_m \end{bmatrix}$$

It is assumed that the measurement vector should contain some error with the exact measurement function value. Therefore, **z** can be written as:

$$\mathbf{z} = \begin{bmatrix} z_1 \\ z_2 \\ \vdots \\ z_m \end{bmatrix} = \begin{bmatrix} h_{1(x_1,x_2,\dots,x_3)} \\ h_{2(x_1,x_2,\dots,x_3)} \\ \vdots \\ h_{m(x_1,x_2,\dots,x_3)} \end{bmatrix} + \begin{bmatrix} e_1 \\ e_2 \\ \vdots \\ e_m \end{bmatrix} = \mathbf{h}(x) + \mathbf{e} \quad (17.2)$$

where, $\mathbf{h}(x) = \begin{bmatrix} h_{1(x_1,x_2,\dots,x_3)} \\ h_{2(x_1,x_2,\dots,x_3)} \\ \vdots \\ h_{m(x_1,x_2,\dots,x_3)} \end{bmatrix}$, $\mathbf{e} = \begin{bmatrix} e_1 \\ e_2 \\ \vdots \\ e_m \end{bmatrix}$, and, $\mathbf{x} = \begin{bmatrix} x_1 \\ x_2 \\ \vdots \\ x_m \end{bmatrix}$

Here, $\mathbf{h}(x)$ is the calculated function values for the state variables. $\mathbf{x}$ is the *vector* of state variables and $\mathbf{e}$ is the vector of measurement errors.

Generally, $\mathbf{e}$ is a zero-mean Gaussian noise vector where measurement errors are independent. Therefore, $\mathbf{E}(\mathbf{e_i}) = 0$, where i = 1, 2, ..., m. And $\mathbf{E}(\mathbf{e_i e_j}) = 0$ and $Cov(\mathbf{e}) = \mathrm{E}[\mathbf{e e}^T] = \mathrm{R} = \mathrm{diag}\,(\sigma_1^2, \sigma_2^2, \dots, \sigma_m^2)$.

### 17.2.2. Calculation of Measurement Function

$\mathbf{h}(x)$ is the vector of calculated functions. Generally, $\mathbf{h}(.)$ is a set of nonlinear functions of the state variables for AC approximations of the load flow equations whereas it would be a set of linear functions if the load flow equations are formulated considering DC approximations.

For a π-model of any network, the measurement function value can be calculated as follows [6]:

a) Real and reactive power injection at bus i:

$$P_i = v_i \sum_{j \in n_i} v_j \left( G_{ij} \cos \delta_{ij} + B_{ij} \sin \delta_{ij} \right) \quad (17.3)$$

$$Q_i = v_i \sum_{j \in n_i} v_j \left( G_{ij} \sin \delta_{ij} - B_{ij} \cos \delta_{ij} \right) \quad (17.4)$$

b) Real and reactive power flow from bus i to bus j are:

$$P_{ij} = v_i^2 \left( g_{si} + g_{ij} \right) - v_i v_j \left( g_{ij} \cos \delta_{ij} + b_{ij} \sin \delta_{ij} \right) \quad (17.5)$$
$$Q_{ij} = -v_i^2 \left( b_{si} + b_{ij} \right) - v_i v_j \left( g_{ij} \cos \delta_{ij} - b_{ij} \sin \delta_{ij} \right) \quad (17.6)$$

c) Line current flow magnitude

$$I_{ij} = \sqrt{\frac{P_{ij}^2 + Q_{ij}^2}{v_i}} \quad (17.7)$$

where, the symbols have their usual meaning.

To calculate the $\mathbf{h}(x)$ values, any other functions can be used based on the formulation of the network model. For example, a multi-phase power flow model is proposed for state estimation in [7].

### 17.2.3. State Estimation: Formulation and Methodologies

As discussed, state estimation depends on the following equation,

$$\mathbf{z} = \mathbf{h}(x) + \mathbf{e} \quad (17.8)$$

Therefore, state estimation can be formulated as an error minimization problem which is in fact a convex optimization problem described below:

$$x' = \arg\min \sum_{i=1}^{m} W(z_i - h_i(x))^2 \quad (17.9)$$

where, W is the weighting matrix which can represents $W = R^{-1}$. To solve the value $x'$, an iterative approach may be adopted. Some popular techniques are Gauss-Newton method and Newton-Raphson method [8]. Evolutionary algorithm, i.e., swarm intelligence based approaches (Particle Swarm Optimization) have also been used to solve this critical operational problem [9].

### 17.2.4. Bad Data Detection

Generally, it is assumed that the measured data will contain some errors. However, sometimes, measured data is so faulty that it affects the state estimation and inconsistent result occurs. As a result, Bad Data Detection becomes very important to obtain a successful state estimation. Different methodologies are used to detect and identify bad data. Such a widely adopted procedure is 'Largest Normalized Residual (LNR)' method [10]. Once the system states ($x'$) are estimated, then the residual is calculated as following

$$r = z - h(x') \quad (17.10)$$

At least one bad data exists if the value of the residual is less than a predefined threshold, which can be written as follows:

$$\text{Bad data exists if } ||r|| < \tau$$

Some other techniques which are also used in literature are 'The $J(x')$ Performance Index', 'Hypotheses Testing', 'Dormant and Perfect Measurement', 'Identification test', etc. [8].

## 17.3. State Estimation for Distribution Networks

State Estimation of transmission system is a well-established area for real-time monitoring and control of a complex power network. However, traditional techniques and methodologies for transmission style State Estimation do not fit for a low voltage power distribution network. Generally, balanced approximation of the power system is considered in most of the traditional State Estimation techniques, e.g., [6], [8]. Although this assumption of positive sequence network modeling is valid for high voltage transmission network, but does not work well for low voltage distribution system [11]. In reality, power lines are transposed and loads are not balanced in a distribution network. Moreover, there are three, two, and single-phase lines and transformers are both delta and wye connected. As a result, rather than a positive sequence modeling of the network, it is essential to have full multi-phase modeling for accurate simulation of distribution network as mentioned in [12]. Considering a-b-c phase modeling, some early researches on Distribution State Estimation are proposed in [11],[13-14] where these issues are clearly pointed out. Other than this multi-phase property and untransposed phase conductors, distribution network exhibits some other characteristics as below [15], [35]:

 a) Feeders are mostly radial in nature
 b) Distributed loads with a small geographical area
 c) High R/X ratio
 d) Presence of Distributed Generation and no conventional generation
 e) Very low redundancy of measurement units

Due to the distinct features of radial low voltage distribution feeders, Distribution State Estimation is different from the traditional one. Moreover, analysis procedure of this real-time operational tool is very challenging because of the following properties [16]:

 a) Limitation of measurement devices
 b) The pseudo-measurement of load data is obtained from the historical load data which may have very limited accuracy
 c) Significant number of current measurement devices are used

These challenges are increased a lot in a smart grid environment which will be discussed in the next section.

## 17.4. Smart Grid State Estimation

Smart Grid State Estimation needs to face the new requirements and challenges of the future renewable energy based sustainable self-healing intelligent smarter grid. Different new aspects will have significant impact on the Smart Grid State Estimation. Three major aspects have been identified in [5] which will be discussed here briefly:

a) Development of Advanced Measurement Technologies: Generally, measurement data of a power system is obtained through the SCADA network. Traditionally, Remote Terminal Units (RTUs) are used for this purpose. RTU is a microprocessor-controlled electronic device that is responsible to measure network traffic through sensors and to transmit the telemetry data to the Distribution Management System for further processing. These measurements are non-synchronized and obtained too infrequently to understand the system operational characteristics. Especially capturing system dynamics is too difficult [5]. In recent years, Phasor Measurement Units (PMUs) have been adopted widely for better real-time monitoring and control of smart grid. PMUs have several advantages over traditional measurement devices, such as:
   - It captures data more frequently, e.g., 20~60 times per second [16].
   - Measurement data are synchronized as they are sampled according to the Global Positioning System (GPS).
   - Current measurement is also possible to those nodes where PMU is placed
b) New regulatory and pricing issues: In a Smart Grid concept, new regulatory issues are arising. In a consumer-centric electricity market, end users are capable to produce electricity and sell those to the Distribution Network Operator (DNO). There arises the need of dynamic pricing and new regulatory issues. As a result, DNO must have a clear knowledge about the whole distribution network, especially regarding the power flows through all the phases of the utility distribution grid. To obtain an accurate power flows, the role of Smart Grid State Estimation is vital.
c) Demand response and Distributed Energy Resources: In order to fulfil the ever growing load demand, co-generation, distributed generation and storages are being employed in the grid. These devices are making the grid active from its traditional passive manner which will introduce a bi-directional power flow [17]. In order to understand the flow pattern, distribution network needs advanced modelling and analysis capabilities which motives to develop advanced State Estimation tools for Smart Grid.

Realizing the needs for developing accurate and fast State Estimation algorithms, a significant number of research works are going on throughout the world [18-23]. A multi-level State Estimation framework for Smart Grid is proposed in [18] where authors propose a new paradigm based on multi-level communication and computation architecture. At the lowest level, a local State Estimation (LSE) is proposed to deal with the distribution substation and its downstream radial feeders. Computed state variable values are then transmitted through the aid of communication infrastructure to its upper level which is Transmission System Operator (TSO) level. Rather than calculating only from 'raw' data, this time TSO-level SE will get a chance to update, smooth and modify the data from LSE by comparing with the raw measurement data. At the final stage, Regional State Estimation (RSE) will synchronize and purify the data obtained from TSO level SE. Numerical simulation is also carried out to explain and evaluate the working procedure of this multi-level schema in [18].

Signal Processing based approaches have also been used to solve this critical real-time operating problem. One such method is Belief Propagation based method to solve Distribution State Estimation [19]. One major challenge of Distribution State Estimation is limited measurement devices. The method proposed in [19] solves the problem of sparse measurement by addressing Belief Propagation based method for real-time Distribution State Estimation. One more advantage of this method is that it can deal with the renewable energy based distributed power generation sources. The performance of the proposed method is compared with the Electric Power Research Institute's distribution system analysis tool openDSS.

In [20], authors propose a method for Distribution State Estimation with the deployment of PMUs and Smart Meters. Authors suggest that rather than mixing the traditional meter measurements and the measurements obtained from Advanced Metering Infrastructure (AMI) and PMU, a two-step approach can be adopted. The advantage is that the current configurations of EMS software need not to be changed. According to that, a traditional State Estimation is performed at the first step and finally PMU measurements are considered to update and modify the pre-processing data.

Considering renewable energy resources, a method for State Estimation based on evolutionary algorithm is proposed in [24]. The proposed method can consider different practical issues including unbalanced power flows, VAR compensators, Voltage Regulators (VRs), tap changing transformers, etc.

## 17.5. Vulnerabilities of Smart Grid Sate Estimation: A Case study

Recent literature shows that a significant improvement of Smart Grid State Estimation is noticeable in terms of 'accuracy' and 'efficiency'. However, this operational tool is very prone to cyber vulnerabilities as discussed in [25]. As the State Estimation is highly dependent on the measurement data, any intruder can inject 'False Data' in such a way that the system is unable to detect it. Fig.17.2 shows such kind of scenario when State Estimation is under attack. This type of malicious modification of measurement data is known as 'False Data Injection Attack' [25] or 'Data Integrity Attack' [31]. In the following sub-section, a step-by-step procedure of a false data injection attack is explained with example.

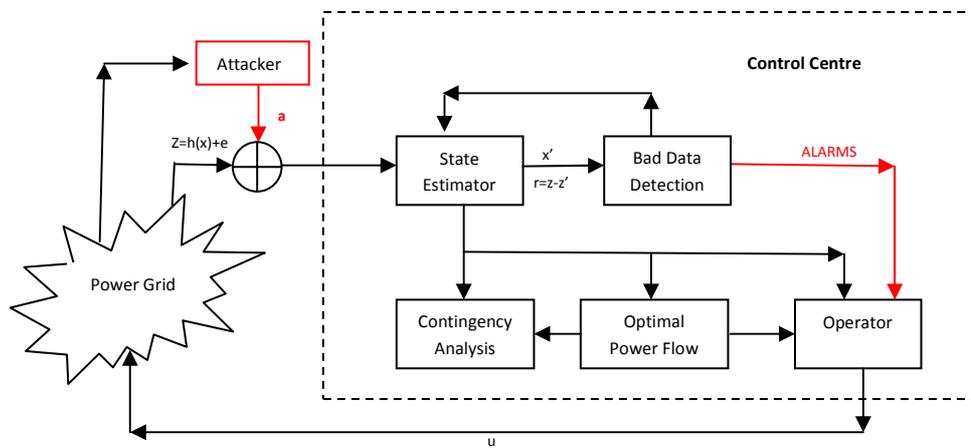

Fig.17.2: State Estimation under attack [34].

In terms of computational complexity, simplified DC approximation of a power network has more advantages over AC model of a power system. Instead of solving 'N' nonlinear equations, one need to solve a set of linear equations in DC approximation where the bus voltage is considered to be known and equivalent to 1 pu. As DC approximation does not need any iterative method, it is faster and reduces the computational burden in the State Estimation process.

For a DC State Estimation, the problem can be defined as:

$$\boldsymbol{z = Hx + e} \qquad (17.11)$$

where, z is the vector of measurement data and $z \in R^N$, $\boldsymbol{H}$ is the Jacobian matrix and $\boldsymbol{e}$ is the error term. When the meter error follows normal distribution with zero mean, the solution becomes as follows [25]:

$$x' = (H^T W H)^{-1} H^T W z \qquad (17.12)$$

where, W is a diagonal matrix as follows:

$$W = \begin{bmatrix} \sigma_1^{-2} & \cdots & \cdot \\ \vdots & \ddots & \vdots \\ \cdot & \cdots & \sigma_m^{-2} \end{bmatrix}$$ and $\sigma_i^{-2}$ is the variance of *i*-th meter.

In order to avoid the bad measurement, the measurement residual $z - Hx$ should be below than the threshold value $\tau$. Generally, it can be said that there is atleast one bad data if $||z - Hx|| > \tau$, otherwise, Bad Data does not exist. However, this assumption is not valid all time. Here, an example is shown to describe how to introduce False Data Injection Attack into the State Estimation. The theoretical concept is adopted from [25].

In this case, a three bus test system is considered as shown in Fig.17.3, where three measurement devices are connected to measure the power through the lines 1-2, 1-3 and 3-2. The measurement powers are $P_{12}$=0.62 pu, $P_{13}$=0.06 pu, $P_{32}$=0.37 pu and $\sigma = 0.01$. Here, $\theta_1$ and $\theta_2$ are the state variables and $\theta_3 = 0$ which is the reference angle. As, the problem solves DC state estimation, the voltages are considered 1 pu.

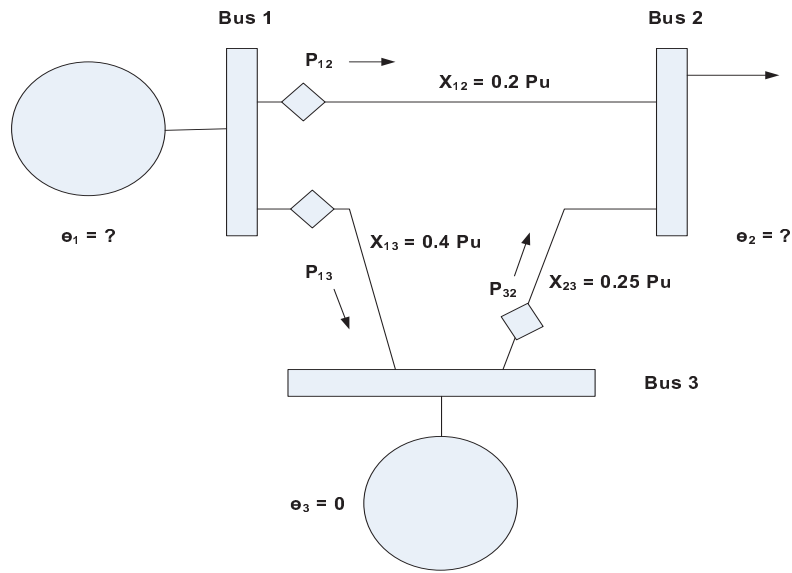

Fig.17.3: Three-bus test system

Following the DC power flow equations:

$$h_1(x) = P_{12} = \frac{(\theta_1-\theta_2)}{X_{12}} = \frac{(\theta_1-\theta_2)}{0.2} = 5(\theta_1 - \theta_2) \tag{17.13}$$

$$h_2(x) = P_{13} = \frac{(\theta_1-\theta_3)}{X_{13}} = \frac{(\theta_1-\theta_3)}{0.4} = 2.5\theta_1 \tag{17.14}$$

$$h_3(x) = P_{32} = \frac{(\theta_3-\theta_2)}{X_{32}} = \frac{(\theta_3-\theta_2)}{0.25} = -4\theta_2 \tag{17.15}$$

So, the H matrix becomes,

$$H = \begin{bmatrix} 5 & -5 \\ 2.5 & 0 \\ 0 & -4 \end{bmatrix}$$

Following the Eqn. (17.12), the values of the state variables become,

$$\theta_1 = 0.0286$$

$$\theta_2 = -0.0943$$

Therefore,

$$\theta = [0.0286 \quad -0.0943]^T$$

Now, the residual matrix *r* becomes,

$$r = (z - Hx') = \begin{bmatrix} 0.62 \\ 0.06 \\ 0.37 \end{bmatrix} - \begin{bmatrix} 5 & -5 \\ 2.5 & 0 \\ 0 & -4 \end{bmatrix} \begin{bmatrix} 0.0286 \\ -0.0943 \end{bmatrix} = \begin{bmatrix} 0.62 \\ 0.06 \\ 0.37 \end{bmatrix} - \begin{bmatrix} 0.614 \\ 0.0714 \\ 0.3771 \end{bmatrix} = \begin{bmatrix} 0.0057 \\ -0.0114 \\ -0.0071 \end{bmatrix}$$

So, the squared error is, $||z - Hx'||^2 = 0.00021429$

This value is very close to zero and it can be said that a good assumption of the state variables are made.

Now, an attack scenario is introduced. It has been assumed that the measurement data is corrupted by the malicious modification of the measured data and therefore $z$ becomes $z_a$ where $z_a = z + a$ and $a$ is the attack vector. Here, $a = (a_1, ..., a_m)^T$ and m is the rank of $z$. It is expected that due to the change of the measured vector the values of the state variables will be altered. Considering that effect, the new state variables will be $x'_{false}$, where $x'_{false} = x' + c$. Here, c is a vector of non-zero values with a length n. According to [25], the residual of the base case ($||z - Hx'||$) and modified case ($||z_a - H x'_{false}||$) would be same if $a = Hc$ that means $a$ is a linear combination of column vectors of H. At this stage four scenarios are considered as described below:

*Scenario 1*: This is the base case where no malicious modification of the measured data is made and therefore, it is assumed that the residual lies within the threshold and there is no probability of a False alarm. In the base case, the values of the measured data are $P_{Base-12}$=0.62 pu , $P_{Base-13}$=0.06 pu, $P_{Base-32}$=0.37 pu.

*Scenario 2*: In this scenario, measured data are modified arbitrarily. Say, the corrupted measurements are $P_{False1-12}$=0.63 pu, $P_{False1-13}$=0.05 pu, $P_{False1-32}$=0.35 pu.

*Scenario 3*: In this scenario, measured data are modified following the Attack Definition proposed in [25]. Considering $c$ as a vector of non-zero arbitrary chosen values of length n:

$c = (c_1, ..., c_n)^T = [0.005 \quad 0.001]^T$

Then, the attack vector $a$ would be

$$a = Hc = \begin{bmatrix} 5 & -5 \\ 2.5 & 0 \\ 0 & -4 \end{bmatrix} * \begin{bmatrix} 0.005 \\ 0.001 \end{bmatrix} = \begin{bmatrix} 0.02 \\ 0.0125 \\ -0.004 \end{bmatrix}$$

So, corrupted measurement would be $z_a = z + a$, therefore,

$P_{False2-12}$= 0.6400 pu , $P_{False2-13}$=0.0725 pu, $P_{False2-32}$= 0.3660 pu.

*Scenario 4:* The attack formulation in this scenario is the same as it is discussed in the previous scenario, however, the value of c is different which is $c = (c_1, ..., c_n)^T = [0.01 \quad 0.04]^T$ )

Therefore,

$$a = Hc = \begin{bmatrix} 5 & -5 \\ 2.5 & 0 \\ 0 & -4 \end{bmatrix} * \begin{bmatrix} 0.01 \\ 0.04 \end{bmatrix} = \begin{bmatrix} -0.15 \\ 0.0250 \\ -0.16 \end{bmatrix}$$

So, corrupted measurement would be $z_a = z + a$, therefore,

$P_{False2-12}$= 0.4700 pu, $P_{False2-13}$= 0.0850 pu, $P_{False2-32}$= 0.2100 pu.

Now, the results obtained from the previous scenarios are discussed. The base case is already discussed in the previous section. According to that, the state variables obtained in the base case are $\theta = [0.0286 \quad -0.0943]^T$. Using these values, the squared error becomes, $||z - Hx'||^2 = 0.00021429$. Considering this error value as normal operating limit ( that means, it is less than the threshold $\tau$ ), we will evaluate how other corrupted measurement data in scenario 2, 3 and 4 perform during the bad data detection in the State Estimation process.

Now, the second scenario is considered where the vector of corrupted measurement data is

$$z_a = \begin{bmatrix} 0.63 \\ 0.05 \\ 0.35 \end{bmatrix}$$

Considering that, the value of $\theta$ becomes,

$$\theta_{Scenario2} = [0.0313 \quad -0.0919]^T$$

So, the residual matrix $r$ becomes,

$$r_{Scenario2} = (z - Hx') = \begin{bmatrix} 0.63 \\ 0.05 \\ 0.35 \end{bmatrix} - \begin{bmatrix} 5 & -5 \\ 2.5 & 0 \\ 0 & -4 \end{bmatrix} \begin{bmatrix} 0.0313 \\ -0.0919 \end{bmatrix} = \begin{bmatrix} 0.0141 \\ -0.0282 \\ -0.0176 \end{bmatrix}$$

Using these values, the squared error becomes, $||z - Hx'||^2 = 0.0013$ which is greater than the squared error in the base case. As a result, the False Data Injection attack may not be overlooked and be detected in the Bad Data Detection test. So, the intruder may not be successful to plan an attack.

At this stage of the discussion, Scenario 3 and Scenario 4 are considered where measurement data are also corrupted like the Scenario 2 but the attack vectors are created following the method described in [25]. Using those corrupted data State Estimation is performed and the obtained state variables are as follows:

$$\theta_{Scenario3} = [0.0336 \quad -0.0933]^T$$

and

$$\theta_{Scenario4} = [0.0386 \quad -0.0543]^T$$

So, the residuals are

$$r_{Scenario3} = (z - Hx') = \begin{bmatrix} 0.6400 \\ 0.0725 \\ 0.3660 \end{bmatrix} - \begin{bmatrix} 5 & -5 \\ 2.5 & 0 \\ 0 & -4 \end{bmatrix} \begin{bmatrix} 0.0336 \\ -0.0933 \end{bmatrix} = \begin{bmatrix} 0.0057 \\ -0.0114 \\ -0.0071 \end{bmatrix}$$

$$r_{Scenario4} = (z - Hx') = \begin{bmatrix} 0.4700 \\ 0.0850 \\ 0.2100 \end{bmatrix} - \begin{bmatrix} 5 & -5 \\ 2.5 & 0 \\ 0 & -4 \end{bmatrix} \begin{bmatrix} 0.0386 \\ -0.0543 \end{bmatrix} = \begin{bmatrix} 0.0057 \\ -0.0114 \\ -0.0071 \end{bmatrix}$$

So, the squared error for both the case is $||z - Hx'||^2 = 0.00021429$.

In scenario 3 and scenario 4, the three measurement devices show different measurement data as they are attacked by the intruder and therefore, the system operator obtains two different set of state variables for these two different scenarios. But it is interesting to note that the residual values and squared errors calculated from both of these scenarios are the same and that is equal to the base case. Therefore, it is expected that the error value is below than the threshold and it will pass the Bad Data Detection technique although attack has been launched. So, the results can be summarized as follows:

Table 17.1: Comparison of different attack scenarios

| Case | False Data Injection Attack | State Variables | | Squared Error | Bad Data Detection |
|---|---|---|---|---|---|
| | | $\theta_1$ | $\theta_2$ | | |
| Scenario 1 | No | 0.0286 | −0.0943 | 0.00021429 | Not Detected |
| Scenario 2 | Yes | 0.0313 | −0.0919 | 0.0013 | Detected |
| Scenario 3 | Yes | 0.0336 | −0.0933 | 0.00021429 | Not Detected |
| Scenario 4 | Yes | 0.0386 | −0.0543 | 0.00021429 | Not Detected |

From the Table 17.1, it can be seen that no bad data is detected in Scenario 1, Scenario 3 and Scenario 4. Although there is no False Data Injection attack in Scenario 1 but measurement data is manipulated in the remaining two scenarios. However, bad data detection technique fails to detect that. It is also interesting to note that squared error term for the above discussed cases is 0.00021429 but state variables vary a lot. As a result, the system operator may take misleading decisions.

## 17.6. Vulnerabilities of Smart Grid State Estimation: A Review

The new class of False Data Injection attack is first proposed in [25]. In that literature, authors show that the DC State Estimation is very vulnerable to malicious modification of the measurement data. With theorem and proof, some heuristic approaches are proposed to attack the DC state estimation. Both random attacks and targeted attacks are considered. Two limitations of the proposed methodology are:

(1) The attacker needs the system 'Configuration' information prior to the attack,

(2) The proposed methodology is developed only for DC State Estimation.

A technique to detect false data injection is proposed in [26], where authors have focused on detecting a set of sensors whose measurements need to be protected in order to capture the false data injection in a DC State Estimator. The work presented in that paper considers the proposed attack model of [25]. The relation between the change of topology and the attack scenario is not considered in this research work [26].

The impact of False Data injection attack on the energy market is discussed in [27] which show that a successful attack can introduce a financial disaster. In that research, a convex optimization problem is formulated to find profitable attack. Although False Data Injection Attack is discussed based on financial issues, the work does not provide any intrusion detection or prevention technique to mitigate the problem.

A protection strategy against the False Data Injection Attack is proposed in [28]. In this research, the authors propose an effective algorithm to identify and protect the key measurements easily. A strategic plan for placement of PMU units is also described here. This work focuses system operator's point of view to utilize limited resources against the False Data Injection Attack. However, this paper does not consider accurate nonlinear AC State Estimation to define and protect attacks in power grid.

Generally, the bad data detection technique relies on the residual errors of the State Estimation procedure. However, it has been proven that this type of methodology for Bad Data Detection is vulnerable to False Data Injection Attacks [25]. To detect bad data, a Generalized Likelihood Ratio Test (GLRT) is proposed in [29]. This paper also considers the False Data Injection Attack from an intruder's point of view where attacker knows the information of mean-square error and GLRT of the system operator. This paper also limits the research idea to DC State Estimation.

Impact of cyber attack on the State Estimation considering a non-linear model is analyzed in [30]. Two widely used Bad Data Detection techniques are considered for comparision. This work concludes that False Data Injection Attack has a better probability to remain undetected if the attacker has a more accurate model of the system.

Another defense strategy against False Data Injection attack is proposed in [32]. The proposed detection framework has two stages. At the first stage, a linear unknown parameter solver is used and finally, a CUSUM algorithm is used to detect the intrusion maintain a certain low level of detection error rate.

Vulnerabilities of AC State Estimation due to the False Data Injection Attack are discussed in [33]. This work extends the hidden False Data Injection Attack model of [25] from a DC approximation to a non-linear AC model. Here, authors propose a Graph-theory based approach to determine critical measurement components which are vulnerable to cyber-attacks.

From the literature review, some decisions may be taken:
> From the attacker's point of view:

a) Detail system model should be considered during the attack creation

b) Different techniques exist to detect the Bad Data. It is important to note that attack vectors should be able to hide against most of the Bad Data Detection techniques.

c) Attack should be introduced with limited knowledge of system and resources.

> From the system operator's point of view:

a) System operator should be aware about the possible attack scenarios.

b) Strategic protection and defense model should be introduced.

A brief description of different types of cyber attacks considering smart grid is given in [36].

## 17.7. Concluding Remarks

The role of State Estimation is crucial to operate the system in a stable condition. In recent time, Smart Grid State Estimation is very vulnerable to False Data Injection Attacks. In this chapter, the overview of State Estimation in both transmission level and distribution level is discussed. The evolution of State Estimation in the Smart Grid and its requirements are also explained. The review of False Data Injection attack is explained with a case study. It is expected that utilities, industries and academics should be more concerned to develop the countermeasures and protection strategies against this type of attacks.

## References

1. R. McMillan, "Siemens: Stuxnet worm hit industrial systems," COMPUTERWorld, Sept.14, 2010.
2. The Industrial Control Systems Cyber Emergency Response Team (ICS-CERT), 'INCIDENT RESPONSE ACTIVITY (April-June 2013)', online available (on 30/08/2013) :http://ics-cert.us-rt.gov/sites/default/files/ICS-CERT_Monitor_April-June2013.pdf
3. Anwar, A.; Pota, H.R., "Optimum allocation and sizing of DG unit for efficiency enhancement of distribution system," *2012 IEEE International Power Engineering and Optimization Conference (PEOCO), Melaka, Malaysia*, vol., no., pp.165,170, 6-7 June 2012
4. Schweppe, F.C.; Wildes, J., "Power System Static-State Estimation, Part I: Exact Model," *IEEE Transactions on Power Apparatus and System*, vol.PAS-89, no.1, pp.120,125, Jan. 1970
5. Yih-Fang Huang; Werner, S.; Jing Huang; Kashyap, N.; Gupta, V., "State Estimation in Electric Power Grids: Meeting New Challenges Presented by the Requirements of the Future Grid," IEEE *Signal Processing Magazine*, vol.29, no.5, pp.33,43, Sept. 2012
6. Ali Abur, Antonio Gómez Expósito, "Power System State Estimation: Theory and Implementation," CRC Press, 2004.
7. Sakis Meliopoulos, A.P.; Fan Zhang, "Multiphase power flow and state estimation for power distribution systems," *IEEE Transactions on Power Systems,* vol.11, no.2, pp.939,946, May 1996
8. Monticelli, A., "Electric power system state estimation," *Proceedings of the IEEE* , vol.88, no.2, pp.262,282, Feb. 2000
9. Naka, S.; Genji, T.; Yura, T.; Fukuyama, Y., "A hybrid particle swarm optimization for distribution state estimation," *IEEE Transactions on Power Systems*, vol.18, no.1, pp.60,68, Feb 2003
10. Handschin, E.; Schweppe, F.C.; Kohlas, J.; Fiechter, A., "Bad data analysis for power system state estimation," *IEEE Transactions on Power Apparatus and System*, vol.94, no.2, pp.329,337, Mar 1975
11. Lu, C.N.; Teng, J.H.; Liu, W.-H.E., "Distribution system state estimation," *IEEE Transactions on Power Systems,* vol.10, no.1, pp.229,240, Feb 1995
12. Kersting, W.H., "The Whys of Distribution System Analysis," *IEEE Industry Applications Magazine*, vol.17, no.5, pp.59,65, Sept.-Oct. 2011
13. Baran, M.E.; Kelley, A.W., "State estimation for real-time monitoring of distribution systems," *IEEE Transactions on Power Systems*, vol.9, no.3, pp.1601,1609, Aug 1994
14. Whei-Min Lin; Jen-Hao Teng, "State estimation for distribution systems with zero-injection constraints," *IEEE Transactions on Power Systems*, vol.11, no.1, pp.518,524, Feb 1996
15. Haughton, D.A.; Heydt, G.T., "A Linear State Estimation Formulation for Smart Distribution Systems," *IEEE Transactions on Power Systems*, vol.28, no.2, pp.1187,1195, May 2013
16. Novosel D., Vu K., Benefits of PMU Technology for Various Applications, *7-th CIGRE Symposium on Power System Management*, Cavtat, Croatia, 5-8 Nov. 2006, pp. 1-13.
17. Anwar, A.; Pota, H.R., "Loss reduction of power distribution network using optimum size and location of distributed generation," *21st Australasian Universities Power Engineering Conference (AUPEC)*, vol., no., pp.1,6, 25-28 Sept. 2011
18. Gomez-Exposito, A.; Abur, A.; de la Villa Jaen, A.; Gomez-Quiles, C., "A Multilevel State Estimation Paradigm for Smart Grids," *Proceedings of the IEEE* , vol.99, no.6, pp.952,976, June 2011

19. Ying Hu; Kuh, A.; Yang, Tao; Kavcic, A., "A Belief Propagation Based Power Distribution System State Estimator," *IEEE Computational Intelligence Magazine*, vol.6, no.3, pp.36,46, Aug. 2011
20. Junqi Liu; Junjie Tang; Ponci, F.; Monti, A.; Muscas, C.; Pegoraro, P.A., "Trade-Offs in PMU Deployment for State Estimation in Active Distribution Grids," *IEEE Transactions on Smart Grid*, vol.3, no.2, pp.915,924, June 2012
21. Sungmin Park; Eunjae Lee; Wonkun Yu; Heungjae Lee; Jeonghoon Shin, "State Estimation for Supervisory Monitoring of Substations," *IEEE Transactions on Smart Grid,* vol.4, no.1, pp.406,410, March 2013
22. Le Xie; Dae-Hyun Choi; Kar, S.; Poor, H.V., "Fully Distributed State Estimation for Wide-Area Monitoring Systems," *IEEE Transactions on Smart Grid,* vol.3, no.3, pp.1154,1169, Sept. 2012
23. Zonouz, S.; Rogers, K.M.; Berthier, R.; Bobba, R.B.; Sanders, W.H.; Overbye, T.J., "SCPSE: Security-Oriented Cyber-Physical State Estimation for Power Grid Critical Infrastructures," *IEEE Transactions on Smart Grid,* vol.3, no.4, pp.1790,1799, Dec. 2012
24. Taher Niknam, Bahman Bahmani Firouzi, A practical algorithm for distribution state estimation including renewable energy sources, *Renewable Energy*, Volume 34, Issue 11, November 2009
25. Y. Liu, M. K. Reiter, and P. Ning, "False data injection attacks against state estimation in electric power grids," *Proceedings of the 16th ACM Conference on Computer and Communications Security*, 2009.
26. BOBBA, R. B., ROGERS, K. M., WANG, Q., and KHURANA, H. 2010. Detecting false data injection attacks on DC state estimation. In Proceedings of the *First Workshop on Secure Control Systems.*
27. Le Xie; Yilin Mo; Sinopoli, B., "False Data Injection Attacks in Electricity Markets," *Smart Grid Communications (SmartGridComm), 2010 First IEEE International Conference on* , vol., no., pp.226,231, 4-6 Oct. 2010
28. Kim, T.T.; Poor, H.V., "Strategic Protection Against Data Injection Attacks on Power Grids," *IEEE Transactions on Smart Grid,* vol.2, no.2, pp.326,333, June 2011
29. Kosut, O.; Liyan Jia; Thomas, R.J.; Lang Tong, "On malicious data attacks on power system state estimation," *45th International Universities Power Engineering Conference (UPEC),* vol., no., pp.1,6, Aug. 31 2010-Sept. 3 2010.
30. Teixeira, A.; Amin, S.; Sandberg, H.; Johansson, K.H.; Sastry, S.S., "Cyber security analysis of state estimators in electric power systems," *49th IEEE Conference on Decision and Control (CDC),* vol., no., pp.5991,5998, 15-17 Dec. 2010
31. Le Xie; Yilin Mo; Sinopoli, B., "Integrity Data Attacks in Power Market Operations," *IEEE Transactions on Smart Grid,* vol.2, no.4, pp.659,666, Dec. 2011
32. Yi Huang; Husheng Li; Campbell, K.A.; Zhu Han, "Defending false data injection attack on smart grid network using adaptive CUSUM test," *45th Annual Conference on Information Sciences and Systems (CISS),* vol., no., pp.1,6, 23-25 March 2011
33. Hug, G.; Giampapa, J.A., "Vulnerability Assessment of AC State Estimation With Respect to False Data Injection Cyber-Attacks," *IEEE Transactions on Smart Grid,* vol.3, no.3, pp.1362,1370, Sept. 2012
34. Teixeira, A.; Amin, S.; Sandberg, H.; Johansson, K.H.; Sastry, S.S., "Cyber security analysis of state estimators in electric power systems," *49th IEEE Conference on Decision and Control (CDC),* vol., no., pp.5991,5998, 15-17 Dec. 2010
35. Anwar, A.; Pota, H.R., "Optimum capacity allocation of DG units based on unbalanced three-phase optimal power flow," *IEEE Power and Energy Society General Meeting*, vol., no., pp.1,8, 22-26 July 2012
36. Anwar, A.; Mahmood, A.N., "Cyber Security of Smart Grid Infrastructure," in *The State of the Art in Intrusion Prevention and Detection,* CRC Press, USA, Jan 2014